\documentclass[epjST]{svjour}
\usepackage{graphicx}
\usepackage[normalem]{ulem}
\usepackage{color}
\usepackage{bm}
\usepackage{amsmath}
\usepackage{xspace}
\usepackage{units}
\usepackage{multicol}

\newcommand{\tenmer}{10-PPE\xspace}
\newcommand{\ppes}{poly {\em para} phenylene ethynylenes\xspace}
\newcommand{\ppe}{poly {\em para} phenylene ethynylene\xspace}

\newcommand{\equ}[1]{eq.~\ref{equ:#1}}
\newcommand{\tab}[1]{Tab.~\ref{tab:#1}}

\newcommand{\sect}[1]{sec.~\ref{sec:#1}}

\newcommand{\Fig}[1]{Figure~\ref{fig:#1}}

\newcommand{\vecr}{{\ensuremath{\mathbf{r}}}}
\usepackage{units}

\begin{document}
\title{Solvent effects on optical excitations of \ppe studied by QM/MM simulations based on Many-Body Green's Functions Theory}
\author{Behnaz Bagheri \and Mikko Karttunen\thanks{\email{m.e.j.karttunen@tue.nl}} \and Bj\"orn Baumeier\thanks{\email{b.baumeier@tue.nl}}}
\institute{Department of Mathematics and Computer Science \& Institute for Complex Molecular Systems, Eindhoven University of Technology, P.O. Box 513, 5600 MB Eindhoven, The Netherlands}
\abstract{
Electronic excitations in dilute solutions of \ppe (poly-PPE) are studied using a QM/MM approach combining many-body Green's functions theory within the $GW$ approximation and the Bethe-Salpeter equation with polarizable force field models. 
Oligomers up to a length of \unit[7.5]{nm} (10 repeat units) functionalized with nonyl side chains are solvated in toluene and water, respectively. After equilibration using atomistic molecular dynamics (MD), the system is partitioned into a quantum region (backbone) embedded into a classical (side chains and solvent) environment. Optical absorption properties are calculated solving the coupled QM/MM system self-consistently and special attention is paid to the effects of solvents. The model allows to differentiate the influence of oligomer conformation induced by the solvation from electronic effects related to local electric fields and polarization. It is found that the electronic environment contributions are negligible compared to the conformational dynamics of the conjugated PPE. An analysis of the electron-hole wave function reveals a sensitivity of energy and localization characteristics of the excited states to bends in the global conformation of the oligomer rather than to the relative of phenyl rings along the backbone.
} 
\maketitle

\section{Introduction}
\label{sec:intro}
\begin{figure}
\centering
\resizebox{\columnwidth}{!}{
\includegraphics[width=\linewidth]{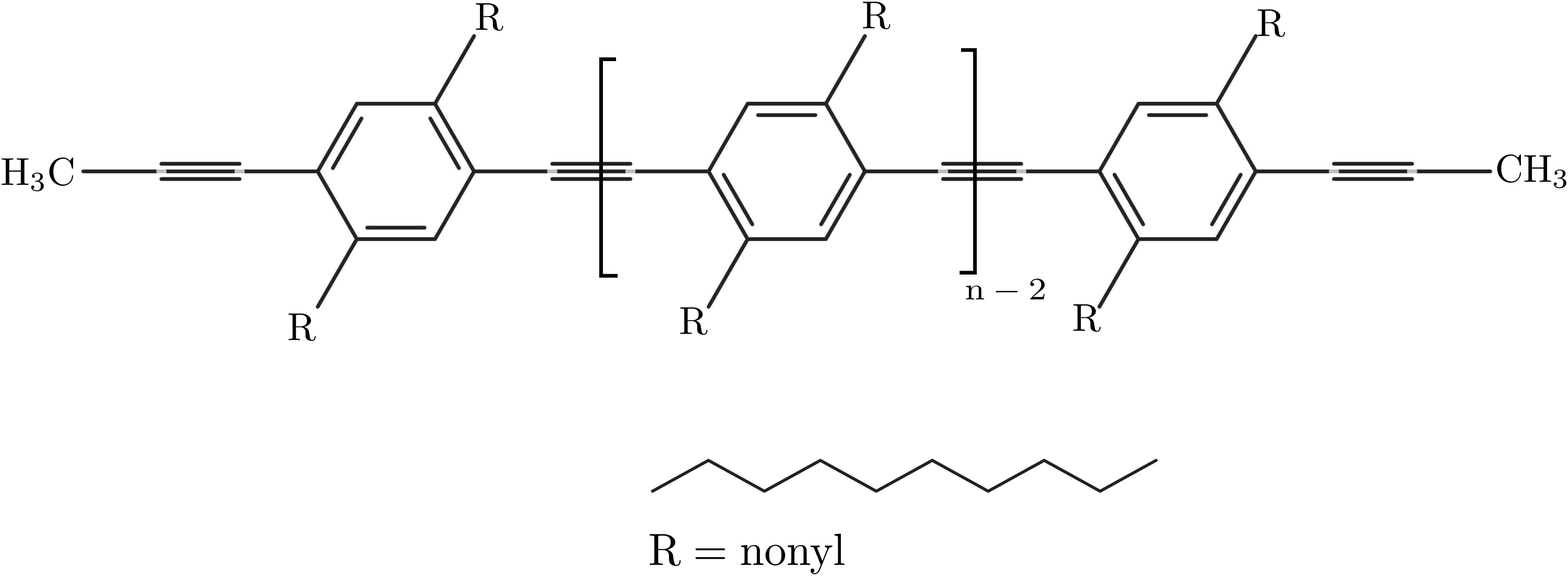}}
\caption{Chemical structure of \ppe (poly-PPE). In this work $R$ indicates nonyl chains. $n$ is the number of repeat units, or the degree of polymerization.}
\label{fig:PPE-chemical-structure}      
\end{figure}

Conjugated polymers have attracted a lot of interest due to their variable functional properties and application potential, e.g., in biochemical sensors~\cite{gaylord_dna_2002,kushon_detection_2002,harrison_amplified_2000,mcquade_conjugated_2000,kim_sensing_2005,liu_fluorescence_2005}, lasers~\cite{hide_semiconducting_1996}, light emitting diodes (LEDs)~\cite{ho_molecular-scale_2000,zhang_light-emitting_1993}, organic transistors~\cite{kaur_solvation_2007}, and photovoltaic devices~\cite{shaheen_organic-based_2005,brabec_polymerfullerene_2010}. The ease of processability and band gap tunability of polymeric semiconductors facilitates the realization of this potential, since it provides the opportunity for a targeted manipulation of electronic and morphological properties of single polymer chains and their aggregates. This, in turn, can be achieved by synthetic strategies, exploitation of properties of functional side chains, and/or solvent-induced transitions. One specific example in which both morphological and electro-optical properties of polymers are purposefully modified is the formation of highly fluorescent conjugated polymer dots for fluorescence imaging in live cells~\cite{wu_multicolor_2008}. Less toxicity together with  flexibility and biocompatibility make these {\em polydots} attractive substitutes for their inorganic counterparts~\cite{halkyard_evidence_1998,tuncel_conjugated_2010}. 

Among fluorescent polymers, \ppes (poly-PPE) (see the chemical structure in~\Fig{PPE-chemical-structure}) are a class of strongly conjugated polymers with a rigid backbone and absorption and emission of light tunable from the ultraviolet (absorption) to the visible (emission) range~\cite{halkyard_evidence_1998}. In particular, PPEs can be used as fluorescence sensors because their fluorescence intensity is sensitive to the presence of other co-solutes. Due to the significance of polymer conformations for the photophysical properties of functionalized PPEs, it is desirable to steer the tendency of the polymer to stay in single strands or to aggregate in a particular solvent by rational manipulations~\cite{yue_evolution_2008}. It has been demonstrated that modification of side chain functionalization by, e.g., substituting surfactants with varying degree of hydrophobicity and solute-solvent or solvent-air interactions, can lead to controlled morphologies and, concomitantly, optical properties. In flexible and semi-flexible polymers conjugation along a single chain may be broken~\cite{vukmirovi_charge_2008,vukmirovi_charge_2009,mcmahon_ad_2009} due to a substantial out-of-plane torsion angle between two adjacent repeat units. However, such a simple criterion may be insufficient to describe and interpret the complex interplay between the actual chemistry of the backbone, functionalization by side chains, and solute-solvent interactions and their cumulative effect on the localization characteristics of excitations and hence the electronic and optical properties of the polymer. 

In this study, a combined quantum-classical (QM/MM) approach is used to investigate optical properties of 2,5-dinonyl \ppe oligomers in dilute solutions with toluene and water, respectively, as model systems. Conformations of the PPE chains are explored by using MD simulations. Electronic excitations are calculated based on many Body Green's Functions theory within the $GW$ approximation and the Bethe-Salpeter equation ($GW$-BSE). The use of the latter technique in traditional quantum-chemical applications has recently increased~\cite{blase_first-principles_2011,baumeier_frenkel_2012,marom_benchmark_2012,van_setten_gw-method_2013}, not least due to its accurate prediction of both localized Frenkel and bimolecular charge transfer excitations~\cite{blase_charge-transfer_2011,baumeier_excited_2012,baumeier_electronic_2014}. Linking $GW$-BSE to a classical environment, represented at atomistic resolution by a polarizable force field, allows for the determination of optical properties in realistic environments from the self-consistent solution of the coupled QM/MM system. With this approach, it is possible to disentangle the conformational (as a result of side chain-solvent interactions) and electronic (due to local electric fields and polarization effects) contributions to the absorption spectra.

The rest of this paper is organized as follows: The methodologies and computational details, are described in \sect{methodology}. The resulting optical properties for isolated oligomers with up to ten repeat units and their sensitivity to structural details are presented in \sect{results_vacuum} and structural properties of dinonyl-10-PPE in solution are discussed in~\sect{results_structure}. In \sect{results_qmmm}, the respective optical absorption spectra resulting from $GW$-BSE/MM calculations are discussed with a focus on the electronic contributions of the environment and the conformational dynamics. A brief summary in \sect{summary} concludes the paper.

\section{Methodology}
\label{sec:methodology}

Classical molecular dynamics (MM/MD) simulations were performed using an OPLS type (optimized potentials for liquid simulations)~\cite{jorgensen_optimized_1984,jorgensen_development_1996,watkins_perfluoroalkanes:_2001} force field. The parameters were taken from Refs.~\cite{maskey_conformational_2011,maskey_internal_2013} in which PCFF (polymer consistent force field) force field parameters were converted to OPLS form. Modification for torsional potential parameters of the phenylene rings as in Ref.~\cite{bagheri_getting_2016} were employed. To study the behavior of PPE polymers in explicit solvents, water molecules were described using the SPC/E~\cite{berendsen_missing_1987} model for water and the OPLS force field for toluene.
Geometric mixing rules [$\sigma_{ij}=(\sigma_{ii}\sigma_{jj})^{\frac{1}{2}}$ and $\epsilon_{ij}=(\epsilon_{ii}\epsilon_{jj})^{\frac{1}{2}}$] for  Lennard-Jones (LJ) diameters ($\sigma$) and LJ energies ($\epsilon$) were used for atoms of different species according to the  OPLS conventions~\cite{jorgensen_optimized_1984,jorgensen_development_1996,watkins_perfluoroalkanes:_2001}. The reader is referred to Ref.~\cite{wong-ekkabut_good_2016} for more in-depth discussion on mixing rules. Non-bonded interactions between atom pairs within a molecule separated by one or two bonds were excluded. Interaction was reduced by a factor of $1/2$ for atoms separated by three bonds and more. Simulations were run using GROMACS version 5~\cite{van_der_spoel_gromacs:_2005}. A \unit[1.2]{nm} cutoff was employed for the real space part of electrostatics and Lennard-Jones interactions. The long-range electrostatics was calculated using particle-mesh Ewald (PME)~\cite{darden_particle_1993,essmann_smooth_1995} with the reciprocal-space interactions evaluated on a 0.16 grid with cubic interpolation of order 4. The importance of proper treatment of electrostatics in MM/MD simulations is discussed in detail in Ref.~\cite{cisneros_classical_2014}. The velocity-Verlet algorithm~\cite{verlet_computer_1967} was employed to integrate the equations of motions with \unit[1]{fs} time step. A Langevin thermostat~\cite{grest_molecular_1986} with a \unit[100]{fs} damping was used to keep the temperature of the system at \unit[300]{K}. The systems  were energy minimized using the steepest descents algorithm. \unit[100]{ps} simulations in constant particle number, volume and temperature (NVT) ensemble at \unit[300]{K} were performed on the energy minimized systems. The simulation box size was $\unit[(15\times 13\times 13)]{nm^3}$ for dinonyl-\tenmer. Simulations were continued in constant particle number, pressure and temperature (NPT) ensemble at \unit[300]{K} and \unit[1]{bar} controlled by Parrinello-Rahman~\cite{parrinello_polymorphic_1981} barostat with a coupling time constant of \unit[2.0]{ps}. Molecular visualizations were done using Visual Molecular Dynamics (VMD) software~\cite{humphrey_vmd:_1996}. 

The excited state electronic structure is evaluated using many-body Green's functions theory within the $GW$ approximation and the Bethe-Salpeter equation (BSE). Green's functions equations of motion are the basis for this approach which contains both the nonlocal energy-dependent electronic self-energy $\Sigma$ and the electron-hole interaction leading to the formation of excitons, described by the BSE. The first of three steps of the procedure is the determination of molecular orbitals and energies on the level of density-functional theory (DFT) by solving the Kohn-Sham equations
\begin{equation}
\left\{ -\frac{\hbar^2}{2m}\nabla^2 + V_\text{PP}(\vecr) + V_H(\vecr) +V_\text{xc}(\vecr)\right\}\psi_n^\text{KS}(\vecr) = \varepsilon_n^\text{KS} \psi_n^\text{KS}(\vecr).
\end{equation}
Here, $V_\text{PP}$ is a pseudo-potential (or effective-core potential), $V_H$ the Hartree potential, and $V_\text{xc}$ the exchange-correlation potential. Single-particle excitations are then obtained within the $GW$ approximation of many-body Green's functions theory, as introduced by Hedin and Lundqvist~\cite{hedin_effects_1969}, by substitution of the energy-dependent self-energy operator $\Sigma(\vecr,\vecr',E)$ for the DFT exchange-correlation potential, giving rise to the quasi-particle equations
\begin{equation}
  \left\{ -\frac{\hbar^2}{2m}\nabla^2 + V_\text{PP}(\vecr) + V_H(\vecr)\right\}\psi_n^\text{QP}(\vecr) + \int{\Sigma(\vecr,\vecr',\varepsilon_n^\text{QP})\psi_n^\text{QP}(\vecr')d\vecr'} = \varepsilon_n^\text{QP} \psi_n^\text{QP}(\vecr).
\end{equation}
The self-energy operator is evaluated as
\begin{equation}
  \Sigma(\vecr,\vecr',E) = \frac{i}{2\pi} \int{e^{-i\omega 0^+}G(\vecr,\vecr',E-\omega)W(\vecr,\vecr',\omega)\,d\omega} ,
\end{equation}
where 
\begin{equation}
  G(\vecr,\vecr',\omega) = \sum_n{\frac{\psi_n(\vecr)\psi_n^*(\vecr')}{\omega-\varepsilon_n+i0^+\text{sgn}(\varepsilon_n -\mu)}}
\label{equ:Green}	
\end{equation}
is the one-body Green's function in quasiparticle (QP) approximation and $W= \epsilon^{-1} v$ is the dynamically screened Coulomb interaction, comprising the dielectric function $\epsilon$, computed within the random-phase approximation, and the bare Coulomb interaction $v$. The ground state Kohn-Sham wave functions and energies are used to determine both $G$ and $W$. Since DFT underestimates the fundamental HOMO-LUMO gap, the self-energy and the resulting QP energies may deviate from self-consistent results. To avoid such deviations, we employ an iterative procedure in which $W$ is calculated from a scissor-shifted Kohn-Sham spectrum. From the resulting QP gap, a new value for the scissor-shift is determined and this procedure is repeated until convergence is reached. For each step, the QP energy levels are iterated and the Green's function of \equ{Green} and thus the self-energy are updated. A one-shot $G_0W_0$ calculation from Kohn-Sham energies may differ from iterated results by up to several 0.1 eV. Note that this (limited) self-consistency treatment does not change the QP structure of \equ{Green} (due to satellite structures or other consequences of a self-consistent spectral shape of $G(\omega)$).

An electron-hole excitation, e.g., resulting from photoexcitation, can not be described in an effective single-particle picture but requires explicit treatment of a coupled two-particle system. Within the Tamm-Dancoff approximation (TDA)~\cite{note_TDA}, the electron-hole wavefunction is given by $\Phi_S(\vecr_e,\vecr_h) = \sum_\alpha^\text{occ}{\sum_\beta^\text{virt}{ A^S_{\alpha\beta}\psi_\beta(\vecr_e)\psi^*_\alpha(\vecr_h)}}$
where $\alpha$ and $\beta$ denote the single-particle occupied and virtual orbitals, respectively, and $A_{\alpha\beta}$ represent the electron-hole amplitudes. These amplitudes and the associated transition energies $\Omega_S$ can be obtained by solving the Bethe-Salpeter equation 
\begin{equation}
(\varepsilon_\beta - \varepsilon_\alpha)A^S_{\alpha\beta} +\sum_{\alpha'\beta'}{K^\text{eh}_{\alpha\beta,\alpha'\beta'}(\Omega_S)A^S_{\alpha'\beta'} } = \Omega_S A^S_{\alpha\beta}
\end{equation}
in which $K^\text{eh}=\eta K^{x}+K^{d}$ ($\eta=2$ for singlets, $\eta=0$ for triplets) is the electron-hole interaction kernel comprised of bare exchange ($K^x$) and screened direct terms ($K^{d})$, respectively. 

For practical calculations according to the $GW$-BSE method, first single-point Kohn-Sham calculations are performed using ORCA~\cite{neese_orca_2012}, the B3LYP functional~\cite{becke_density-functional_1993,lee_development_1988,vosko_accurate_1980,stephens_ab_1994}, effective core potentials of the Stuttgart/Dresden type~\cite{bergner_ab_1993}, and the associated basis sets that are augmented by additional polarization functions~\cite{krishnan_self-consistent_1980} of $d$ symmetry~\cite{note_ECP}. All steps involving the actual $GW$-BSE calculations are performed using the implementation for isolated systems~\cite{ma_excited_2009,ma_modeling_2010,baumeier_excited_2012,baumeier_frenkel_2012}, available in the VOTCA software package~\cite{ruhle_microscopic_2011,note_VOTCA}. 
In VOTCA, the quantities in the $GW$ self-energy operator (dielectric matrix, exchange and correlation terms) and the electron-hole interaction in the BSE are expressed in terms of auxiliary atom-centered Gaussian basis functions. We include orbitals of $s$, $p$, $d$ symmetry with the decay constants $\alpha$ (in a.u.) 0.25, 0.90, 3.0 for C and 0.4 and 1.5 for H atoms, yielding converged excitation energies. It was also confirmed that the addition of diffuse functions with decay constants smaller than \unit[0.06]{a.u.} to the wave function basis set does not affect the low-lying excitations. For all systems considered in this paper, polarizability is calculated using the full manifold of occupied and virtual states in the random-phase approximation. Quasiparticle corrections are calculated for the $2n_\text{occ}$ lowest-energy states, and $n_\text{occ}$ occupied and $n_\text{occ}$ virtual states are considered in the Bethe-Salpeter equation. Further technical details can be found in Refs~\cite{ma_excited_2009,ma_modeling_2010,baumeier_excited_2012}.

\section{Results}
\label{sec:results}

\subsection{Optical absorption energies of isolated oligomers}
\label{sec:results_vacuum}
\begin{table}\centering
\caption{Electronic structure data for $n$-PPE oligomers with $n=1,\dots,10$ based on QM and MM optimized geometries: HOMO-LUMO gap from Kohn-Sham ($E_g^\text{KS}$), quasi-particle ($E_g^\text{QP}$) energies, optical excitation energy ($\Omega$), and the contributions to it from free inter-level transitions ($\langle D \rangle$) and electron-hole interaction ($\langle K^\text{eh} \rangle = \langle K^d + 2K^x\rangle$). All energies in eV. }
\label{tab:energies}       
\begin{tabular}{lccccccccccc}
\hline\noalign{\smallskip}
& \multicolumn{5}{c}{\bf{QM optimized}} & & \multicolumn{5}{c}{\bf{MM optimized}} \\
\cline{2-6}  \cline{7-12}\noalign{\smallskip}
$n$ & $E_g^\text{KS}$ & $E_g^\text{QP}$ & $\Omega$ & $\langle D \rangle$ & $\langle K^\text{eh} \rangle$ & & $E_g^\text{KS}$ & $E_g^\text{QP}$ & $\Omega$ & $\langle D \rangle$ & $\langle K^\text{eh} \rangle$  \\
\noalign{\smallskip}\hline\noalign{\smallskip}
1 & 4.67 & 8.46 & 5.15 & 9.11 & -3.96 & & 4.71 & 8.52 & 5.15 & 9.14 & -3.99 \\
2 & 3.80 & 7.07 & 4.17 & 7.49 & -3.32 & & 3.96 & 7.30 & 4.28 & 7.72 & -3.44 \\
3 & 3.42 & 6.45 & 3.73 & 6.84 & -3.11 & & 3.61 & 6.73 & 3.89 & 7.13 & -3.24 \\
4 & 3.22 & 6.12 & 3.50 & 6.51 & -3.01 & & 3.45 & 6.46 & 3.71 & 6.88 & -3.17 \\
5 & 3.10 & 5.92 & 3.36 & 6.32 & -2.96 & & 3.33 & 6.26 & 3.57 & 6.70 & -3.13 \\
6 & 3.02 & 5.79 & 3.27 & 6.21 & -2.94 & & 3.25 & 6.13 & 3.48 & 6.59 & -3.11 \\
7 & 2.97 & 5.69 & 3.21 & 6.13 & -2.92 & & 3.19 & 6.04 & 3.43 & 6.51 & -3.08 \\
8 & 2.94 & 5.63 & 3.17 & 6.09 & -2.92 & & 3.18 & 6.01 & 3.41 & 6.50 & -3.07 \\
9 & 2.90 & 5.58 & 3.13 & 6.04 & -2.91 & & 3.17 & 5.98 & 3.39 & 6.47 & -3.08 \\
10 & 2.88 & 5.53 & 3.11 & 6.01 & -2.90 & & 3.15 & 5.95 & 3.37 & 6.46 & -3.09 \\
$\infty$ & & & 3.08 &  & & & & &  3.33 & &  \\ 
Exp. & & \multicolumn{3}{c}{3.00 - 3.20 } \\
\noalign{\smallskip}\hline
\end{tabular}
\end{table}

Conformations of solvated PPE oligomers as obtained from classical MD simulations form the basis for the determination of optical excitations in a mixed QM/MM setup. The underlying assumption is that the details of molecular geometries resulting from the use of a classical force field are consistent with  the chosen quantum mechanical description of the ground state. To confirm this the geometries of single $n$-PPE oligomers with $n=1,\dots,10$ were optimized in vacuum using both DFT (def2-TZVP~\cite{weigend_balanced_2005} basis set and B3LYP functional with Grimme's D3 dispersion corrections~\cite{grimme_semiempirical_2006}) and MD on the basis of the modified PCFF force field. Both approaches yield planar configurations of the PPE backbone for all values of $n$. 

While qualitatively identical, quantitative differences can be observed. Most notably, the bonds in the phenyl rings and the $C-C$ bond connecting the ring to the ethyne are elongated by about 2\% in MD compared to DFT. In contrast, the length of the $C\equiv C$ triple bond results \unit[1.21]{\AA} in both cases. In the next step, $GW$-BSE calculations were performed to gauge the effect of these differences on the electronic and optical properties of the oligomers. The resulting electronic structure data, summarized in \tab{energies}, illustrates the effects of $GW$-BSE with respect to the calculation of excitation energies: Taking the quasi-particle corrections to the Kohn-Sham energies into account increases the HOMO-LUMO gap for, e.g., the QM optimized \tenmer from \unit[2.88]{eV} to \unit[5.53]{eV}, which reflects the well-known underestimation of the fundamental gap by DFT. The energy for the lowest optically active coupled electron-hole excitation gives \unit[3.11]{eV}. Due to the fact that the excitation is not a pure HOMO-LUMO transition but has additional  contributions from lower occupied and higher virtual single-particle orbitals, the contribution of the independent transitions, $\langle D \rangle$, is with \unit[6.01]{eV} slightly larger than $E_g^\text{QP}$. The associated effectively attractive electron-hole interaction, $\langle K^\text{eh} \rangle = \langle K^d + 2K^x\rangle$, in this structure amounts to \unit[2.90]{eV}. The obtained excitation energy is in good agreement with the experimental values of \unit[3.0-3.2]{eV} obtained from absorption peaks of dilute solutions of PPE in good solvents. As a function of the number of repeat units $n$, a monotonous decrease of all energies is found. This quantum-size effect is anticipated for strongly conjugated systems such as PPEs. From the particle-in-a-box model one can estimate, e.g., the optical excitation energy of an infinitely long chain via $\Omega(n) = \Omega_\infty - a/n$. By fitting the data for $n>3$ to this model, one obtains a value of $\Omega_\infty = \unit[3.08]{eV}$ indicating that for  further studies of solvent effects on the excitations in PPE, it is reasonable to consider \tenmer in the QM/MM setup. 

Qualitatively identical results were obtained for the $GW$-BSE calculations based on the MM optimized geometries. However, there are some noticeable quantitative differences, most importantly with respect to the excitation energy $\Omega$ which is consistently larger as compared to the QM structures. For large $n$, the deviation amounts to \unit[0.25]{eV} and is a cumulative result of the geometric differences discussed above. Overall, the agreement is satisfying enough to conclude that the use of MD simulated conformations of \tenmer in the following is capturing the relevant physico-chemical details.

\subsection{Structural properties of solvated 2,5-dinonyl-\tenmer}
\label{sec:results_structure}
\begin{figure}
\centering
\includegraphics[width=\linewidth]{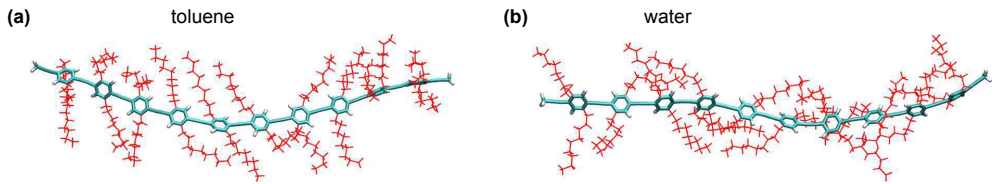}
\caption{Structures of 2,5-dinonyl-\tenmer (a) in toluene and (b) in  water after \unit[7.7]{ns} MD simulations in the NpT ensemble. (a): In toluene, the side chains are dispersed and separated from each other as well as the backbone. (b): In water, the side chains start to aggregate toward the backbone. }
\label{fig:toluene_water_structures}      
\end{figure}

\begin{figure}
\centering
\resizebox{0.70\columnwidth}{!}{
\includegraphics[width=\linewidth]{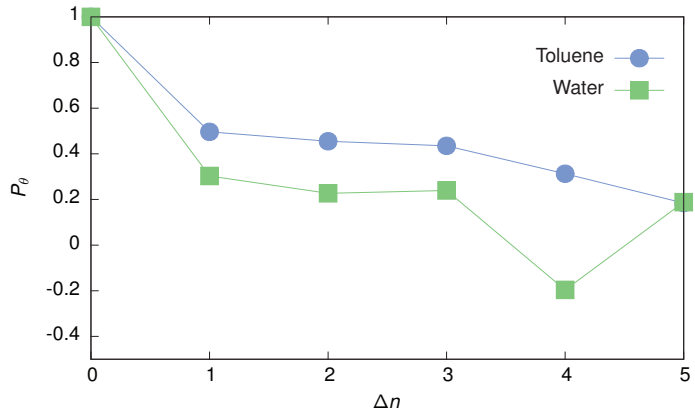}}
\caption{Orientations order parameter (Eq.~\ref{equ:orderparameterformula}) for \tenmer with nonyl side chains in toluene solvent (blue circles) and water (green squares). The length  of MM/MD simulations for both was 7.7\,ns in the NpT ensemble. Time average for toluene case is taken over frames of the last 1\,ns trajectory with 100\,ps  between the frames. In the case of water, the time average was taken over frames of the last 600\,ps of the trajectory with a 100\,ps step.}
\label{fig:OrderParameter}       
\end{figure}

Conformations of 2,5-dinonyl-\tenmer were studied in explicit water and toluene. Water is a poor solvent for both the backbone and the side chains while toluene is a good solvent for the backbone and a poor solvent for the side chains~\cite{maskey_conformational_2011,maskey_internal_2013}. Figure~\ref{fig:toluene_water_structures} shows the structure of \tenmer (a) in toluene and (b) in water at 7.7\,ns in the NpT ensemble. For clarity, water and toluene molecules are not shown. In toluene, the backbone remains extended and the side chains are dispersed and separated from each other as well as the backbone. This is in agreement with the results of Ref.~\cite{maskey_conformational_2011}. Structural studies using small angle neutron scattering (SANS) have shown that dialkyl PPE forms a molecular solution with an extended backbone at high temperature and low concentrations~\cite{maskey_conformational_2011,perahia_molecules_2001}. In water (Fig.~\ref{fig:toluene_water_structures}(b)), the side chains start to aggregate toward each other and the backbone. This is in agreement with Ref.~\cite{maskey_conformational_2011,maskey_internal_2013}. Another important parameter is the correlation of aromatic rings along the backbone of PPE. The interplay between the arrangements of aromatic rings in PPE polymers and their electro-optical properties has been studied by several groups (see e.g.~\cite{bunz_polyaryleneethynylenes_2009,miteva_interplay_2000,kong_molecular_2015}).

The orientational order parameter\cite{hariharan_structure_1994}, given by 
\begin{equation}
P_{\theta}=\frac{1}{2}\langle 3 \cos^2\theta -1\rangle
\label{equ:orderparameterformula}
\end{equation}
is a measure to quantify how aromatic rings within PPE polymer backbone are correlated. $\theta$ is the angle between the normal vectors to the planes of two aromatic rings which are apart from each other by a distance $\Delta n$. $P_{\theta}$ describes the average alignment of aromatic rings. Since each vector normal to the planes of the aromatic rings can be considered as a reference direction to calculate $\theta$ and then $P_{\theta}$, one needs to consider the vector normal to each plane as a reference direction and take an average: there are two averages in the calculation of $P_{\theta}$, one of which is the time average over time frames and the other one is over the selection of a vector normal to the plane of the rings. $P_{\theta}$ can have values [$-\frac{1}{2}$,1]~\cite{hariharan_structure_1994}. $P_{\theta}>0$ describes a co-planar alignment of aromatic rings, while $P_{\theta}<0$ indicates perpendicular alignments. $P_{\theta}=0$ and $P_{\theta}=1$ refer to completely random and fully co-planar alignment of the rings, respectively~\cite{maskey_conformational_2011,maskey_internal_2013}. 
Figure~\ref{fig:OrderParameter} shows the order parameter versus $\Delta n$ for \tenmer with nonyl side chains in toluene and water after~\unit[7.7]{ns} MD simulations in the NpT ensemble. The time average is taken over the frames of the last \unit[1]{ns} (\unit[0.6]{ns}) of the MM/MD trajectory  with 100\,ps time step between the frames for \tenmer in toluene (water).  Having a value of around 0.4 (0.2) in toluene (water) indicates a correlation between the aromatic rings. This refers to an angle of around $39^{\circ}$ and $47^{\circ}$ for \tenmer in toluene and water, respectively. In Ref.~\cite{bunz_polyaryleneethynylenes_2009}, the authors discussed optical properties of dialkyl and dialkoxy-PPEs  in chloroform and dichloromethane, and the average angle of aromatic rings using a  configuration-coordinate model in which  they concluded the angle to be around 40 degrees.

\subsection{Optical absorption of solvated 2,5-dinonyl-10-PPE}
\label{sec:results_qmmm}
Excitation energies in complex molecular environments, such as molecular aggregates or solute-solvent mixtures, can accurately be obtained by treating the subpart of interest quantum-mechanically and embedding it into an environment at molecular mechanics resolution~\cite{risko_quantum-chemical_2011,may_can_2012,lunkenheimer_solvent_2013,schwabe_pericc2:_2012}. Here, we realize such a QM/MM scheme based on $GW$-BSE by representing the molecules in the MM region by a set of atomic properties such as static partial charges and polarizabilities, which then interact among each other and the $GW$-BSE part via classical electrostatic potentials. Polarizable interactions are taken into account via Thole's model~\cite{thole_molecular_1981,van_duijnen_molecular_1998}. For both neutral and excited complexes, total energies of the combined QM/MM system are obtained self-consistently and their difference defines the excitation energy in the polarizable environment. This procedure assumes that the states of interest and, in particular, their localization characteristics on the QM cluster are easily identifiable. Typically, this can be expected to be the case for the lowest optically active excitations in the PPEs studied here.

\begin{figure}
\includegraphics[width=\linewidth]{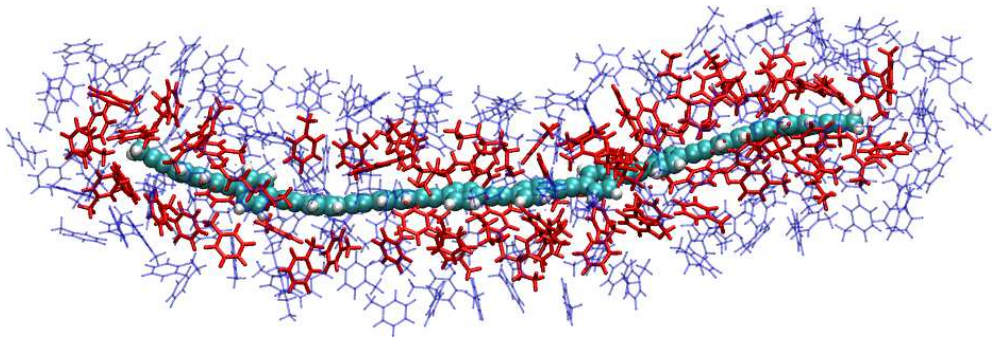}
\caption{Example of $GW$-BSE/MM partitioning of the system. The oligomer is embedded into a two-layer environment of solvent molecules. Molecules within a region $R_1$ (red) are represented by both static atomic point charges and polarizabilities, while then ones within the extended layer $R_2$ (blue) are only represented by point charges.}
\label{fig:qmmm_cutoffs}
\end{figure}

A two-layer scheme is employed. Within a cutoff $R_1$ around the QM part, atomic partial and polarizable interactions are taken into account. In the more extended buffer region with $R_2 \geq R_1$, only electrostatics are active. An example for such a partitioning is depicted in \Fig{qmmm_cutoffs}. For a snapshot taken from the MD simulated morphology of \tenmer with nonyl side chains in toluene, this approach is adopted using cutoffs of $R_1=\unit[2.5]{nm}$ and $R_2=\unit[4.0]{nm}$. The \tenmer backbone is treated quantum-mechanically while the side chains and solvent molecules belong to the MM region. To split the functionalized oligomer into backbone and side chains, a link-atom approach~\cite{reuter_frontier_2000} with hydrogen saturation of the backbone-side chain bridge are employed. Partial charges for the solvent molecules and side chain fragments are determined from CHELPG~\cite{breneman_determining_1990} fits to the electrostatic potentials, while the atomic Thole polarizabilities are parametrized to match the molecular polarizability tensors obtained from DFT. 

\begin{figure}
\includegraphics[width=\linewidth]{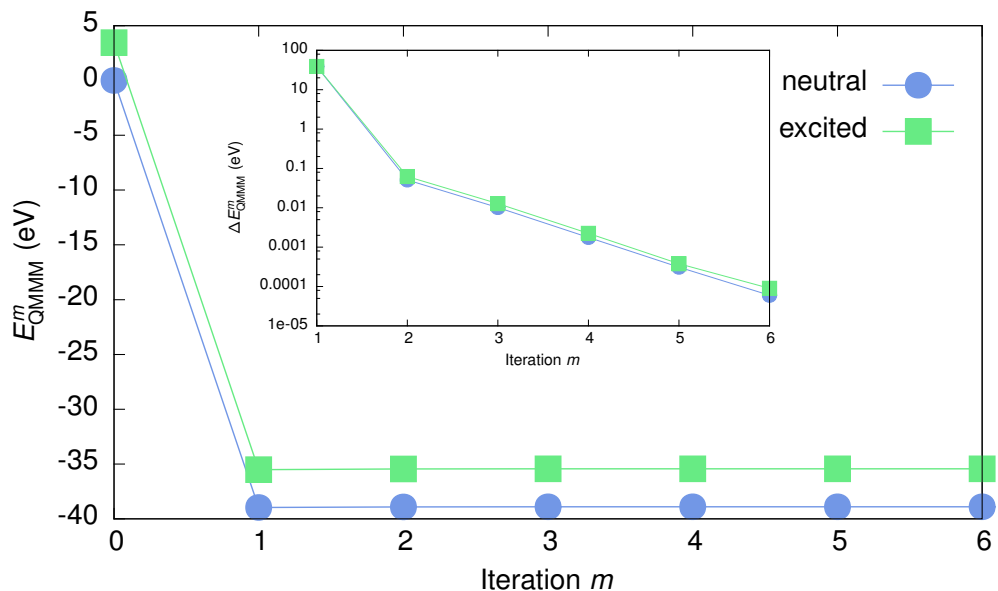}
\caption{Total energies of the coupled QM/MM system ($E_\text{QMMM}^m$) for a representative snapshot at each iteration step $m$ of the self-consistent procedure in neutral and excited states. The inset shows the respective energy differences between two subsequent iteration steps. }
\label{fig:qmmm_convergence}
\end{figure}

\Fig{qmmm_convergence} shows a typical evolution of total energies of the coupled QM/MM system $E_\text{QMMM}$ during the self-consistency procedure~\cite{note_QMMM}. The zero of the energy scale is defined to correspond to the total energy of the neutral \tenmer in vacuum (iteration $m=0$). It is apparent that the most significant change to the total energy in both neutral and excited states occurs during the very first step of the calculation. This is further corroborated by considering the change of total energy at iteration $m$ compared to the previous iteration as shown in the inset of \Fig{qmmm_convergence}. Within three iterations the respective changes are of the order of \unit[0.01]{eV} and, more importantly, no significant differences are observed for the two states. Overall, the effect of polarization is small for the solvated \tenmer and consequently, the excitation energy is nearly unaffected by the environment. More precisely, the excitation energy is \unit[3.47]{eV}   from the polarized QM/MM system, while a calculation using pure static MM yields \unit[3.44]{eV}. Omitting the environment altogether, i.e., performing a $GW$-BSE calculation on the isolated oligomer conformation, yields \unit[3.45]{eV}. The fact that environment effects only have a negligible impact on the calculated excitation energies can be attributed to a combination of the diluteness of the solution and the associated randomness of local electric fields and the small change of dipole moment between neutral and excited states. Similar observations have been made, e.g., for the optically excited states of push-pull oligomers embedded in a polarizable lattice~\cite{baumeier_electronic_2014}.

\begin{figure}
\includegraphics[width=\linewidth]{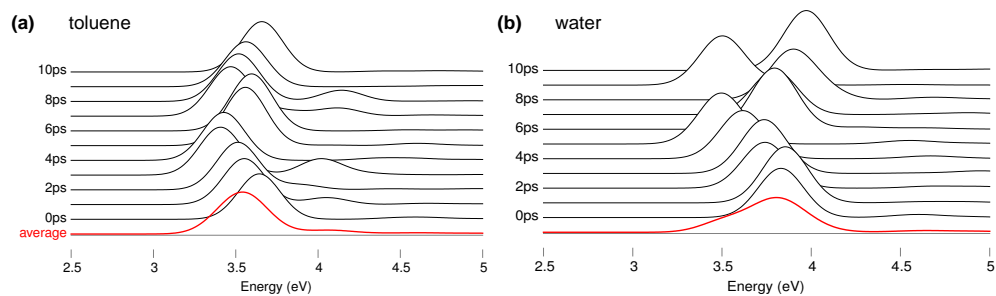}
\caption{Simulated absorption spectra (broadened by Gaussian functions with a FWHM of $\unit[0.3]{eV}$) of \tenmer in (a) toluene and (b) water calculated in a static MM environment ($R_2=\unit[4]{nm}$) with a sampling time step $\Delta t=\unit[1]{ps}$ starting from $t_0=\unit[7.7]{ns}$. The average over the eleven respective snapshots is given in red.  }
\label{fig:spectra_toluene_water}      
\end{figure}

Based on the above results, it is justified to limit the QM/MM setup to only electrostatic interactions in the following. Having realized that the direct electronic effects of solvent molecules on the excitations in \tenmer are small, the focus is now on indirect effects that originate from the influence on the backbone conformations. To this end, \tenmer in both  toluene and water is considered and sample the conformations at different time intervals ($\Delta t = \unit[10]{fs}, \unit[100]{fs}, \unit[1]{ps}$, and $\unit[10]{ps}$), all starting from an identical starting point $t_0=\unit[7.7]{ns}$ of our MD simulations are taken. For each of these snapshots the absorption spectrum is calculated in a static MM environment defined by $R_2=\unit[4]{nm}$ (implies $R_1=\unit[0]{nm}$). The obtained discrete spectra of excitation energies and associated oscillator strengths are broadened by Gaussian functions with a FWHM (full width at half maximum) of \unit[0.3]{eV}. It is found that the absorption properties are insensitive to structural dynamics of the backbone at time scales of \unit[100]{fs}. Only for times exceeding about \unit[500]{fs} fluctuations in the peak positions and heights of the spectra can be observed both in toluene and water. \Fig{spectra_toluene_water} shows the evolution of the absorption spectrum for the time steps $\Delta t = \unit[1]{ps}$, as well as the average over the eleven respective snapshots. While the dynamics of the backbone is comparatively slow since it is to a significant extent constrained by the nonyl side chain dynamics in both poor solvents, one can observe stronger fluctuations of the absorption spectra in water as compared to toluene.



\begin{figure}
\includegraphics[width=\linewidth]{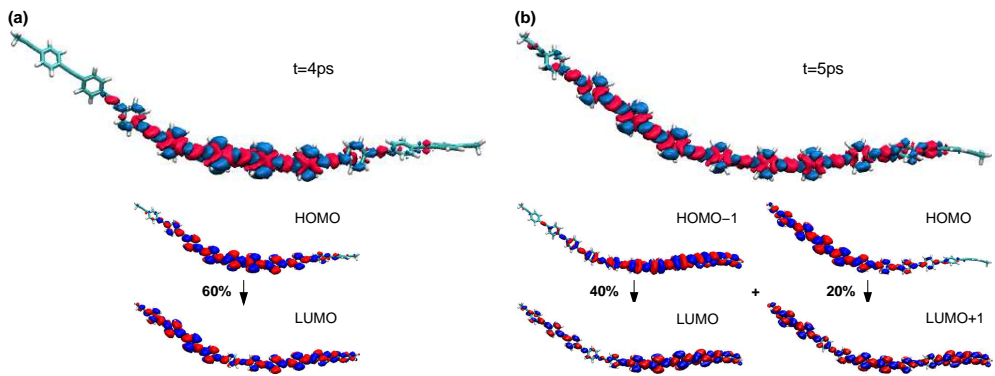}
\caption{Analysis of the excited state wave functions for representative snapshots of 10-PPE in toluene. 
Top row: Isosurfaces of excitation electron density ($\pm10^{-4}\unit{e/{\AA}^3}$). 
Red color corresponds to negative values (hole density), blue to positive values (electron density). 
Bottom rows: Isosurfaces of the main single-particle excitations contributing to the electron-hole 
wave functions (isovalue $\pm 5\cdot10^{-3}$).}
\label{fig:excitations}
\end{figure}

To understand the origin of these fluctuations with respect to backbone conformations in more detail, the electron-hole wave functions of the excitations is analyzed at times $t=\unit[4]{ps}$ and $t=\unit[5]{ps}$, c.f. \Fig{spectra_toluene_water}(a). In the top row of \Fig{excitations} isosurfaces for the hole (red) and electron (blue) density distributions are shown. The overall conformation of the \tenmer exhibits a characteristic bend as a result of the stress caused by side chain interactions.  At both times, the excitation appears to be localized at the apex of the bend, more pronounced for the structure at \unit[4]{ps} which is lower in energy by \unit[0.13]{eV}. The different characteristics can be attributed to a slightly stronger out-of-plane bent angle between the phenylene and etynlene. Over all, co-planarity of the phenyl rings along the backbone (or the lack thereof) does not appear to affect the excitations significantly.

Analysis of the composition of the electron-hole wavefunction reveals striking differences between the two snapshots. The excitation shown in \Fig{excitations}(a) is formed to 60\% by a transition between the two frontier orbitals. The isosurfaces of these orbitals show that both HOMO and LUMO are practically extended along the full length of the backbone. Slight intensity variations can be noted with the HOMO being more attracted to the apex while the LUMO is thinning out at the same spot. These variations give rise to the coupled excitation being localized. At $t=\unit[5]{ps}$, in contrast, there is not a single dominant contribution to the electron-hole wavefunction. Rather, a superposition of several transitions is found, with HOMO-1 to LUMO and HOMO to LUMO+1 transitions being most significant. As can be seen in \Fig{excitations}(b) conformational changes result in a different localization characteristics of the underlying single-particle orbitals not at the apex but left and right from it, respectively. As a pure transition between two localized states, such as the one from HOMO-1 to LUMO, is energetically penalized by stronger exchange interactions. By mixing in transitions between lower lying occupied and higher unoccupied levels, an effectively more delocalized excitation is formed. An analogous analysis of the respective excitations of 10-PPE in water, i.e, at $t=\unit[9]{ps}$ and $t=\unit[10]{ps}$, reveals qualitatively similar behaviour.

\section{Summary}
\label{sec:summary}
Electronic excitations of PPE were computed using a QM/MM approach combining many-body Green's functions theory within $GW$ approximation and the Bethe-Salpeter equation.  Conformations of solvated PPE as obtained from atomistic MD simulations were used in the mixed QM/MM setup in order to determine optical excitations of solvated  PPE. The reliability of optical excitations based on MM/MD conformations were investigated by comparing optical excitations of $n$-PPE ($n=1,2,\dots,10$) using both optimized DFT geometries and MD geometries in vacuum. The results show that the excitation energies $\Omega$ calculated based on MM/MD conformations are larger than the ones calculated based on QM optimized geometries. For large $n$, the deviation amounts to \unit[0.25]{eV} and is the cumulative result of geometric differences between MM/MD geometries and QM geometries. Overall agreement between the excitation energies  based on MM/MD conformations and QM geometries is good enough  to conclude that the use of MM/MD conformations for \tenmer  captures the relevant physico-chemical properties.        

Conformations of 2,5-dinonyl-\tenmer with nonyl side chains were studied in toluene and water. The side chains were found to be dispersed from each other and from the backbone in toluene. In water, the side chains tend to  aggregate.  
Optical excitations were calculated for \tenmer in the QM/MM setup. The results show that the electronic environment contributions are negligible compared to the conformation dynamics of the conjugated PPE. From the analysis of the electron-hole wave function, sensitivity of energy and localization characteristics of the excited states to bends in global conformation of PPE polymer was observed.


\begin{thebibliography}{72}

\bibitem{gaylord_dna_2002}
B.S. Gaylord, A.J. Heeger, G.C. Bazan, Proceedings of the National Academy of
  Sciences \textbf{99}(17), 10954 (2002)

\bibitem{kushon_detection_2002}
S.A. Kushon, K.D. Ley, K.~Bradford, R.M. Jones, D.~McBranch, D.~Whitten,
  Langmuir \textbf{18}(20), 7245 (2002)

\bibitem{harrison_amplified_2000}
B.S. Harrison, M.B. Ramey, J.R. Reynolds, K.S. Schanze, Journal of the American
  Chemical Society \textbf{122}(35), 8561 (2000)

\bibitem{mcquade_conjugated_2000}
D.T. McQuade, A.E. Pullen, T.M. Swager, Chemical Reviews \textbf{100}(7), 2537
  (2000)

\bibitem{kim_sensing_2005}
I.B. Kim, A.~Dunkhorst, J.~Gilbert, U.H.F. Bunz, Macromolecules
  \textbf{38}(11), 4560 (2005)

\bibitem{liu_fluorescence_2005}
M.~Liu, P.~Kaur, D.H. Waldeck, C.~Xue, H.~Liu, Langmuir \textbf{21}(5), 1687
  (2005)

\bibitem{hide_semiconducting_1996}
F.~Hide, M.A. Diaz-Garcia, B.J. Schwartz, M.R. Andersson, Q.~Pei, A.J. Heeger,
  Science \textbf{273}(5283), 1833 (1996)

\bibitem{ho_molecular-scale_2000}
P.K.H. Ho, J.S. Kim, J.H. Burroughes, H.~Becker, S.F.Y. Li, T.M. Brown,
  F.~Cacialli, R.H. Friend, Nature \textbf{404}(6777), 481 (2000)

\bibitem{zhang_light-emitting_1993}
C.~Zhang, D.~Braun, A.J. Heeger, Journal of Applied Physics \textbf{73}(10),
  5177 (1993)

\bibitem{kaur_solvation_2007}
P.~Kaur, H.~Yue, M.~Wu, M.~Liu, J.~Treece, D.H. Waldeck, C.~Xue, H.~Liu, The
  Journal of Physical Chemistry B \textbf{111}(29), 8589 (2007)

\bibitem{shaheen_organic-based_2005}
S.~Shaheen, D.~Ginley, G.~Jabbour, MRS bulletin \textbf{30}(1), 10 (2005)

\bibitem{brabec_polymerfullerene_2010}
C.J. Brabec, S.~Gowrisanker, J.J.M. Halls, D.~Laird, S.~Jia, S.P. Williams,
  Advanced Materials \textbf{22}(34), 3839 (2010)

\bibitem{wu_multicolor_2008}
C.~Wu, B.~Bull, C.~Szymanski, K.~Christensen, J.~McNeill, ACS Nano
  \textbf{2}(11), 2415 (2008)

\bibitem{halkyard_evidence_1998}
C.E. Halkyard, M.E. Rampey, L.~Kloppenburg, S.L. Studer-Martinez, U.H.F. Bunz,
  Macromolecules \textbf{31}(25), 8655 (1998)

\bibitem{tuncel_conjugated_2010}
D.~Tuncel, H.V. Demir, Nanoscale \textbf{2}(4), 484 (2010)

\bibitem{yue_evolution_2008}
H.~Yue, M.~Wu, C.~Xue, S.~Velayudham, H.~Liu, D.H. Waldeck, The Journal of
  Physical Chemistry B \textbf{112}(28), 8218 (2008)

\bibitem{vukmirovi_charge_2008}
N.~Vukmirovi\'{c}, L.~Wang, The Journal of Chemical Physics \textbf{128}(12),
  121102 (2008)

\bibitem{vukmirovi_charge_2009}
N.~Vukmirovi\'{c}, L.~Wang, Nano Letters \textbf{9}(12), 3996 (2009)

\bibitem{mcmahon_ad_2009}
D.P. {McMahon}, A.~Troisi, Chemical Physics Letters \textbf{480}(4-6), 210
  (2009)

\bibitem{blase_first-principles_2011}
X.~Blase, C.~Attaccalite, V.~Olevano, Physical Review B \textbf{83}(11), 115103
  (2011)

\bibitem{baumeier_frenkel_2012}
B.~Baumeier, D.~Andrienko, M.~Rohlfing, Journal of Chemical Theory and
  Computation \textbf{8}(8), 2790 (2012)

\bibitem{marom_benchmark_2012}
N.~Marom, F.~Caruso, X.~Ren, O.T. Hofmann, T.~Körzdörfer, J.R. Chelikowsky,
  A.~Rubio, M.~Scheffler, P.~Rinke, Physical Review B \textbf{86}(24), 245127
  (2012)

\bibitem{van_setten_gw-method_2013}
M.J. van Setten, F.~Weigend, F.~Evers, Journal of Chemical Theory and
  Computation \textbf{9}(1), 232 (2013)

\bibitem{blase_charge-transfer_2011}
X.~Blase, C.~Attaccalite, Applied Physics Letters \textbf{99}(17), 171909
  (2011)

\bibitem{baumeier_excited_2012}
B.~Baumeier, D.~Andrienko, Y.~Ma, M.~Rohlfing, J. Chem. Theory Comput.
  \textbf{8}(3), 997 (2012)

\bibitem{baumeier_electronic_2014}
B.~Baumeier, M.~Rohlfing, D.~Andrienko, Journal of Chemical Theory and
  Computation \textbf{10}(8), 3104 (2014)

\bibitem{jorgensen_optimized_1984}
W.L. Jorgensen, J.D. Madura, C.J. Swenson, Journal of the American Chemical
  Society \textbf{106}(22), 6638 (1984)

\bibitem{jorgensen_development_1996}
W.L. Jorgensen, D.S. Maxwell, J.~Tirado-Rives, Journal of the American Chemical
  Society \textbf{118}(45), 11225 (1996)

\bibitem{watkins_perfluoroalkanes:_2001}
E.K. Watkins, W.L. Jorgensen, The Journal of Physical Chemistry A
  \textbf{105}(16), 4118 (2001)

\bibitem{maskey_conformational_2011}
S.~Maskey, F.~Pierce, D.~Perahia, G.S. Grest, The Journal of Chemical Physics
  \textbf{134}(24), 244906 (2011)

\bibitem{maskey_internal_2013}
S.~Maskey, N.C. Osti, D.~Perahia, G.S. Grest, ACS Macro Letters \textbf{2}(8),
  700 (2013)

\bibitem{bagheri_getting_2016}
B.~Bagheri, B.~Baumeier, M.~Karttunen, submitted  (2016)

\bibitem{berendsen_missing_1987}
H.J.C. Berendsen, J.R. Grigera, T.P. Straatsma, The Journal of Physical
  Chemistry \textbf{91}(24), 6269 (1987)

\bibitem{wong-ekkabut_good_2016}
J.~Wong-ekkabut, M.~Karttunen, Biochimica et Biophysica Acta (BBA) -
  Biomembranes  (2016)

\bibitem{van_der_spoel_gromacs:_2005}
D.~Van Der~Spoel, E.~Lindahl, B.~Hess, G.~Groenhof, A.E. Mark, H.J.C.
  Berendsen, Journal of Computational Chemistry \textbf{26}(16), 1701 (2005)

\bibitem{darden_particle_1993}
T.~Darden, D.~York, L.~Pedersen, The Journal of Chemical Physics
  \textbf{98}(12), 10089 (1993)

\bibitem{essmann_smooth_1995}
U.~Essmann, L.~Perera, M.L. Berkowitz, T.~Darden, H.~Lee, L.G. Pedersen, The
  Journal of Chemical Physics \textbf{103}(19), 8577 (1995)

\bibitem{cisneros_classical_2014}
G.A. Cisneros, M.~Karttunen, P.~Ren, C.~Sagui, Chemical Reviews
  \textbf{114}(1), 779 (2014)

\bibitem{verlet_computer_1967}
L.~Verlet, Physical Review \textbf{159}(1), 98 (1967)

\bibitem{grest_molecular_1986}
G.S. Grest, K.~Kremer, Physical Review A \textbf{33}(5), 3628 (1986)

\bibitem{parrinello_polymorphic_1981}
M.~Parrinello, Journal of Applied Physics \textbf{52}(12), 7182 (1981)

\bibitem{humphrey_vmd:_1996}
W.~Humphrey, A.~Dalke, K.~Schulten, Journal of Molecular Graphics
  \textbf{14}(1), 33 (1996)

\bibitem{hedin_effects_1969}
L.~Hedin, S.~Lundqvist, in \emph{Solid {State} {Physics}: {Advances} in
  {Research} and {Application}} (Academix Press, New York, 1969), Vol.~23, pp.
  1--181

\bibitem{note_TDA}
For typical donor molecules used in organic solar cells, we showed that the use
  of the TDA overestimates $\pi$-$\pi$ transition energies by \unit[0.2]{eV}
  but yields correct character of the excitations~\cite{baumeier_excited_2012}.

\bibitem{neese_orca_2012}
F.~Neese, Wiley Interdisciplinary Reviews: Computational Molecular Science
  \textbf{2}(1), 73 (2012)

\bibitem{becke_density-functional_1993}
A.D. Becke, The Journal of Chemical Physics \textbf{98}(7), 5648 (1993)

\bibitem{lee_development_1988}
C.~Lee, W.~Yang, R.G. Parr, Physical Review B \textbf{37}(2), 785 (1988)

\bibitem{vosko_accurate_1980}
S.H. Vosko, L.~Wilk, M.~Nusair, Canadian Journal of Physics \textbf{58}(8),
  1200 (1980)

\bibitem{stephens_ab_1994}
P.J. Stephens, F.J. Devlin, C.F. Chabalowski, M.J. Frisch, The Journal of
  Physical Chemistry \textbf{98}(45), 11623 (1994)

\bibitem{bergner_ab_1993}
A.~Bergner, M.~Dolg, W.~Küchle, H.~Stoll, H.~Preu{\textbackslash}s~s,
  Molecular Physics \textbf{80}, 1431 (1993)

\bibitem{krishnan_self-consistent_1980}
R.~Krishnan, J.S. Binkley, R.~Seeger, J.A. Pople, The Journal of Chemical
  Physics \textbf{72}, 650 (1980)

\bibitem{note_ECP}
The use of ECPs offers a computational advantage as the wave functions entering
  the $GW$ procedure are smooth close to the nuclei and do not require strongly
  localized basis functions, keeping the numerical effort tractable. We
  confirmed that the Kohn-Sham energies obtained from ECP-based calculations do
  not deviate significantly from all-electron results.

\bibitem{ma_excited_2009}
Y.~Ma, M.~Rohlfing, C.~Molteni, Physical Review B \textbf{80}(24), 241405
  (2009)

\bibitem{ma_modeling_2010}
Y.~Ma, M.~Rohlfing, C.~Molteni, Journal of Chemical Theory and Computation
  \textbf{6}(1), 257 (2010)

\bibitem{ruhle_microscopic_2011}
V.~Rühle, A.~Lukyanov, F.~May, M.~Schrader, T.~Vehoff, J.~Kirkpatrick,
  B.~Baumeier, D.~Andrienko, Journal of Chemical Theory and Computation
  \textbf{7}(10), 3335 (2011)

\bibitem{note_VOTCA}
Available on {\tt www.votca.org}.

\bibitem{weigend_balanced_2005}
F.~Weigend, R.~Ahlrichs, Physical Chemistry Chemical Physics \textbf{7}(18),
  3297 (2005)

\bibitem{grimme_semiempirical_2006}
S.~Grimme, Journal of Computational Chemistry \textbf{27}(15), 1787 (2006)

\bibitem{perahia_molecules_2001}
D.~Perahia, R.~Traiphol, U.H.F. Bunz, Macromolecules \textbf{34}(2), 151 (2001)

\bibitem{bunz_polyaryleneethynylenes_2009}
U.H.F. Bunz, Macromolecular Rapid Communications \textbf{30}(9-10), 772 (2009)

\bibitem{miteva_interplay_2000}
T.~Miteva, L.~Palmer, L.~Kloppenburg, D.~Neher, U.H.F. Bunz, Macromolecules
  \textbf{33}(3), 652 (2000)

\bibitem{kong_molecular_2015}
H.~Kong, G.~He, Molecular Simulation \textbf{41}(13), 1060 (2015)

\bibitem{hariharan_structure_1994}
A.~Hariharan, J.G. Harris, The Journal of Chemical Physics \textbf{101}(5),
  4156 (1994)

\bibitem{risko_quantum-chemical_2011}
C.~Risko, M.D. McGehee, J.L. Bredas, Chemical Science \textbf{2}(7), 1200
  (2011)

\bibitem{may_can_2012}
F.~May, B.~Baumeier, C.~Lennartz, D.~Andrienko, Physical Review Letters
  \textbf{109}(13), 136401 (2012)

\bibitem{lunkenheimer_solvent_2013}
B.~Lunkenheimer, A.~Köhn, Journal of Chemical Theory and Computation
  \textbf{9}(2), 977 (2013)

\bibitem{schwabe_pericc2:_2012}
T.~Schwabe, K.~Sneskov, J.M. Haugaard~Olsen, J.~Kongsted, O.~Christiansen,
  C.~Hättig, Journal of Chemical Theory and Computation \textbf{8}(9), 3274
  (2012)

\bibitem{thole_molecular_1981}
B.~Thole, Chemical Physics \textbf{59}(3), 341 (1981)

\bibitem{van_duijnen_molecular_1998}
P.T. van Duijnen, M.~Swart, J. Phys. Chem. A \textbf{102}(14), 2399 (1998)

\bibitem{reuter_frontier_2000}
N.~Reuter, A.~Dejaegere, B.~Maigret, M.~Karplus, The Journal of Physical
  Chemistry A \textbf{104}(8), 1720 (2000)

\bibitem{breneman_determining_1990}
C.M. Breneman, K.B. Wiberg, Journal of Computational Chemistry \textbf{11}(3),
  361 (1990)

\bibitem{note_QMMM}
In the evaluation of the total QM/MM energy, the contribution resulting from
  the interactions of the static MM partial charges is neglected, since we are
  ultimately only interested in total energy differences. Note that we do,
  however, account for the effect of the electric field of these charges on the
  polarizable quantum and classical parts.

\end{thebibliography}

\end{document}